\documentclass[twocolumn,showpacs,preprintnumbers,amsmath,amssymb]{revtex4}
\usepackage{epsfig}

\begin{document}
\draft

\title{One dimensional heat conductivity exponent from random collision model}

\author{J. M. Deutsch and Onuttom Narayan}
\affiliation{
Department of Physics, University of California, Santa Cruz, CA 95064.}

\date{\today}

\begin{abstract}
We study numerically the thermal conductivity coefficient $\kappa$ as a
function of system length $L$ for several different quasi one dimensional
models: classical gases of hard spheres with both longitudinal and
transverse degrees of freedom. We introduce a model that is ergodic
and highly chaotic but also conserves energy and momentum, and is very
useful because it shows scaling even at small system sizes.  We find
that $\kappa \sim L^\alpha$ over more than two decades, with $\alpha$
very close to the analytical prediction of 1/3.

\end{abstract}


\maketitle 

Since the surprising result obtained over thirty years ago that the heat
current flowing across a one-dimensional chain of harmonic oscillators
with a small temperature difference between the two ends is independent
of the length $L$ of the chain~\cite{Lieb}, the conductivity of one
dimensional systems has been studied analytically~\cite{Exact,Dhar,LLP}
and numerically~\cite{LLP,Numeric,Grass,Grass2,Casati} at great
length. The standard approach to conductivity would predict that if
the temperature gradient $\nabla T$ in a material is small, the heat
current flowing through should be of the form $j = - \kappa\nabla T,$
where $\kappa$ is a property of the material. On the other hand, the
result for the harmonic oscillator chain, that a small temperature
difference $\Delta T$ results in a current $j\propto \Delta T$ that is
independent of $L$, is equivalent to a conductivity $\kappa\propto L.$
In the various models studied thereafter, singular conductivities
$\kappa\sim L^\alpha$
have been found, with a variety of possible values for $\alpha.$
On the other hand, for some one-dimensional models a conventional
$L$-independent $\kappa$ has also been obtained.

It is now believed that the singular conductivity of these models has
two possible causes. Firstly, if the model is integrable, as in the
case of the harmonic oscillator chain, the system does not equilibrate
thermally.  The behavior of the conductivity then depends on the
details of the system. In fact, it has been shown that by changing the
coupling of the oscillator chain to the heat reservoirs at the ends,
normally a benign procedure, one can tune the exponent $\alpha$ over
a range~\cite{Dhar}. Secondly, even if the model is not integrable,
if the internal interactions in the system conserve momentum, the
conductivity is singular, due to advection of heat in long wavelength
modes~\cite{ONSR,Grass,Prozen}. If a model is not integrable and does
not conserve momentum, $\kappa$ should have a well defined limit as
$L$ diverges~\cite{dingaling}. Analytical studies~\cite{ONSR,Grass}
predict that for non-integrable momentum conserving systems, $\alpha$
should have a universal value of $1/3.$

Numerical results for momentum conserving systems have yielded values of
$\alpha$ ranging from 0.25 to 0.5. Recent studies of one dimensional
chains of hard point particles with alternating masses have shown
unexpectedly large corrections to scaling even for systems of $\sim
10^4$ particles, with $\alpha$ estimated to be $0.25$~\cite{Casati}
and $0.33$~\cite{Grass}. Similar results have been obtained for
chains of Fermi Pasta Ulam chains, with $\alpha$ estimated to be
0.37~\cite{Grass2}. For the system of hard point particles, the slow
convergence to the asymptotic behavior has been justified~\cite{Grass}
by noticing that the system is always `close' to an integrable model.
Thus if one considers a chain of particles with equal masses, energy is
carried ballistically, the system does not thermalize, and $\alpha = 1.$
On the other hand, if the ratio of the masses of successive particles
in the alternating chain is chosen to be very different from unity,
the light particles are almost inconsequential, and the problem again
reduces to one of equal mass particles. In Ref.~\cite{Grass}, a mass
ratio of 2.62 was found to yield the longest power-law scaling range,
from which $\alpha$ was estimated.  

In this paper, we study how to eliminate residual effects of integrability
by considering non-integrable and highly chaotic models. We start with  
Sinai and Chernov's pencase model~\cite{pencase} that was initially introduced
by them to study one dimensional hydrodynamics. In this model,
hard sphere particles are confined to a long narrow tube. We consider 
both periodic and hard wall boundary conditions in the transverse direction,
and apply heat baths to the two ends. The extent in the transverse
direction is taken to be slightly less than twice the diameter of the
particles. This ensures that the particles cannot get past each other,
but allows a large range of incidence angles at the collisions. Thus the
transport of energy along the tube remains quasi one dimensional, with
the transverse degree of freedom serving as an additional randomizing
effect. This model (without heat reservoirs) has been proved to be
ergodic in four dimensions and hyperbolic in three~\cite{SSz}.

The extra degree of freedom in our model limits the system sizes we
simulate. For the range of system sizes that we consider, there is still
insufficient universality to obtain the asymptotic value for $\alpha$
with confidence. Therefore, we have devised a novel 1d model that we call
the ``random collision" (RC) model, discussed in detail below, that is
extremely chaotic, yet still satisfies energy and momentum conservation.

Probably due to the further randomization introduced by our new dynamics,
the conductivity fits very well to the form $\kappa\sim L^\alpha,$
with $\alpha$ close to 1/3. If the masses of all the particles are
taken to be equal, the estimated value of $\alpha$ is $0.29\pm 0.01,$
while when the particle masses alternate with a mass ratio of 2.62,
one obtains $\alpha$ to be $0.335\pm 0.01.$ The small discrepancy from
the theoretical prediction of $\alpha = 1/3,$ although larger than the
error bars, is within the range one might expect from corrections to
scaling from the leading irrelevant operators.

We also show results for the autocorrelation function of the energy
current, which is related to the conductivity through the Kubo
formula~\cite{Kubo}. The autocorrelation function has a zero frequency
limit that scales close to $L^{1/3},$ in accordance with previous
predictions and our numerical results for the conductivity.  However,
the full correlation function $C(\omega, L)$ shows quite complicated
behavior, not easily described by any simple scaling form, possibly
indicating the existence of multiple time scales. 

One might be concerned whether the analytical derivation of $\alpha =
1/3,$ which uses the hydrodynamic description of a one-dimensional normal
fluid~\cite{ONSR} is valid for the models we have considered here.
Since the transverse direction is small, it should not affect long
wavelength singularities in the dynamics. Of greater concern is, with
periodic boundary conditions, the existence of the transverse momentum
as another conserved quantity.  Although this makes the hydrodynamic
equations more complicated, and increases their number from three to four,
we recall that the analytical calculation relies only on the fact that
(without an applied temperature gradient) the system reaches thermal
equilibrium, that it satisfies Galilean invariance, and that a finite
sound velocity sets a cutoff to the dynamics for any finite system size.
None of these conditions is violated by the introduction of the transverse
momentum.

\begin{figure}
\centerline{\epsfxsize=4in \epsfbox{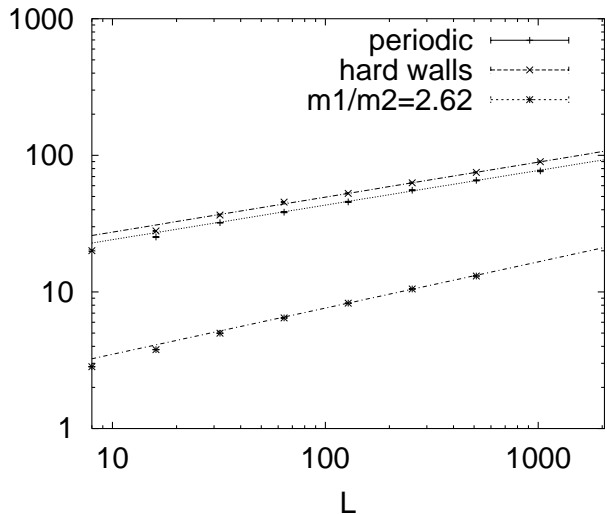}}
\caption{
Log-log plot of the conductivity as a function of the number of particles
for Sinai's pencase model~\cite{pencase}. The upper two plots are for 
periodic and hard wall boundary conditions in the transverse direction,
with the mass of all particles being 1. The slopes of these plots, which
should be equal to the conductivity exponent $\alpha$,  are 
$0.25\pm 0.02$ and $0.26\pm 0.01$ respectively. The lowest plot is for 
hard wall boundary conditions in the transverse direction, with the mass of 
particles alternating between 2.62 and 1. The slope of the plot is 
$0.34\pm 0.02.$ Even though each individual plot fits well to a straight
line (apart for deviations at the low $L$ end), the differences between 
the slopes preclude a good estimate of the asymptotic large $L$ value of 
$\alpha.$
}
\label{2dtube}
\end{figure}
Figure~\ref{2dtube} 
shows a log log plot of the conductivity as a function of the length the
system $L,$ for the pencase model.  The results for both periodic and
hard wall boundary conditions are shown. All the particles were taken
to have the same mass.  The diameter of the particles was 0.6, while the
cross section of the tube and the average longitudinal separation between
centers of neighboring particles were both taken to be 1. The temperatures
of the heat reservoirs at the two ends were taken to be $1.0$ and $1.2$
respectively; it was verified numerically that this temperature difference
is sufficiently small for the system to be in the linear response
regime. The heat reservoirs at the ends were implemented as follows:
whenever an extremal particle collided with the reservoir adjoining it,
its velocity was randomized, drawn from the distribution $P(v_x, v_y)
\propto v_x\exp[-m(v_x^2 + v_y^2)/(2 k_B T)], $ where $x$ and $y$ are
along the longitudinal and transverse direction respectively and $T$
is the temperature of the reservoir. (This is the velocity distribution
for particles leaking out of a heat reservoir.)  At each such collision,
the energy exchanged by the system and the reservoir is kept track of,
and used to calculate the time average of the energy current at both ends.

In the same figure, the conductivity as a function of system length
is also shown for a different choice of model parameters: alternating
particle masses, with a mass ratio of 2.62, a tube cross section of
1.14 and a longitudinal interparticle separation of 0.9. (Only hard wall
transverse boundary conditions are shown for this case.)

The number of particles in the system ranged from 8 to 1024. This is
substantially less than the largest system sizes used when simulating
the one dimensional chain of hard point particles. However, introducing
the transverse dimension should make the system no longer near
integrability, and therefore allow the large $L$ limit to be reached
quickly. Unfortunately, as seen in Figure~\ref{2dtube}, this is not
the case. There is some curvature in all the plots; more importantly,
there is substantial disagreement between the slopes obtained when the
particles have equal or alternating masses. Note that the plots all
curve {\it downwards\/}, from which one might be tempted to conclude
that the asymptotic slope (i.e. the value of $\alpha$) would be smaller
than obtained from the curves. However, prior experience with the one
dimensional system~\cite{Grass} indicates that it is possible for the
curves to turn around at much larger system sizes. Thus one cannot
obtain even an upper bound to $\alpha$ from the figure, and must
conclude that the corrections to scaling are large for
this model~\cite{foot1}.

In order to randomize the dynamics further, enabling faster convergence
to the asymptotic scaling form for large $L,$ we modify the model above.
Firstly, the diameter of the particles is taken to be negligibly
small, while keeping the cross-section of the tube as less than twice
the diameter. Secondly, the particles are no longer disk shaped, but highly
irregular. As a result, when two particles collide with each other, in
the center of mass frame they can recoil in any direction, unrelated to
the direction of impact. For any collision, we take the recoil angle in
the center of mass frame to be a uniform random variable~\cite{foot2}.
The result of both these modifications together is that the transverse
coordinate $y$ of the particles becomes redundant, and they effectively
move along the $x$ axis. However, the transverse velocities $v_y$
are retained.  In this RC model, each particle has both
$v_x$ and $v_y,$ with the latter behaving as an auxiliary variable
that is only important in collisions. In any interparticle collision,
the total momentum in the $x$ and $y$ directions and the total energy
are conserved. Collisions with the reservoirs at the two ends are still
implemented as before. In the transverse direction, periodic
boundary conditions correspond to $v_y$ for a particle remaining
constant between collisions, whereas hard wall boundary conditions allow
$v_y$ to be reversed.
In the latter case, since in the zero cross-section
limit any particle undergoes a huge number of collisions with the
sidewalls between two collisions with its neighbors, one should change
the sign of $v_y$ randomly between interparticle collisions.

\begin{figure}
\centerline{\epsfxsize=4in \epsfbox{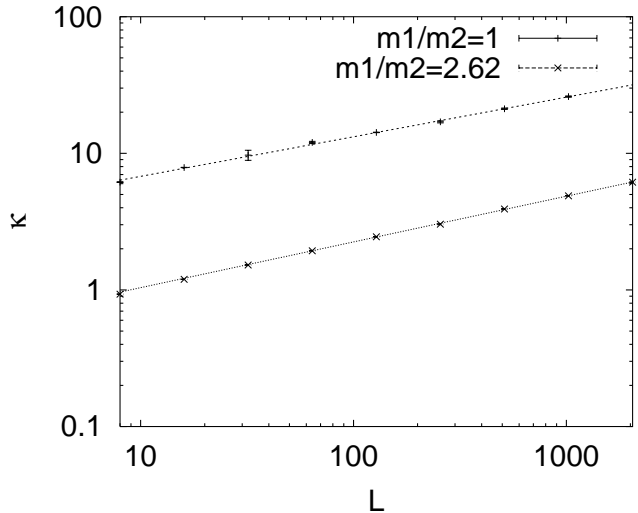}}
\caption{
Log-log plot of the conductivity as a function of the number of particles
for the RC model introduced in this paper. The upper plot 
is for all the particles with mass 1, while the lower plot is for
particles whose masses alternate between 1 and 2.62. The slopes for the 
two plots are $0.29\pm 0.01$ and $0.335\pm 0.01.$ Both the plots are for
periodic boundary conditions in the transverse direction; the results 
for hard wall boundary conditions were very similar. 
}
\label{random}
\end{figure}

Figure~\ref{random} 
shows a log log plot of the conductivity as a function of $L$ for
the RC model. The particles are
now points, the average interparticle separation is unity, and the
temperatures of the reservoirs are 1 and 1.2 as before. In the linear
response regime, the dependence of the energy current on the average
interparticle separation and the reservoir temperatures can be found
trivially, so the only parameters that can be varied (apart from the
number of particles) are the masses of the particles. Only the results
for periodic boundary conditions in the transverse direction are shown,
as the results for the hard-wall case are very similar. As before,
plots are shown for the case when all particles have the same mass
and the case when the particle masses alternate with a mass ratio of
2.62. Both parameter choices yield plots that rapidly converge to a
power law form, but with slightly different exponents. The estimated
exponents are $\alpha = 0.29\pm 0.01$ and $\alpha = 0.335\pm 0.01$
respectively for the two cases. Both of these are very close to the
analytical prediction of $\alpha = 1/3.$ Although the difference between
the two estimates for $\alpha$ is larger than the error bars, we expect
that this is due to corrections to scaling from irrelevant operators in
a renormalization group analysis; an $O(1)$ bare strength for irrelevant
operators can produce effective values of $\alpha$ that differ from $1/3$
by the desired amount. Thus the numerical results of Figure~\ref{random}
are a strong indication of the validity of the prediction of $\alpha =1/3.$

\begin{figure}
\centerline{\epsfxsize=\columnwidth \epsfbox{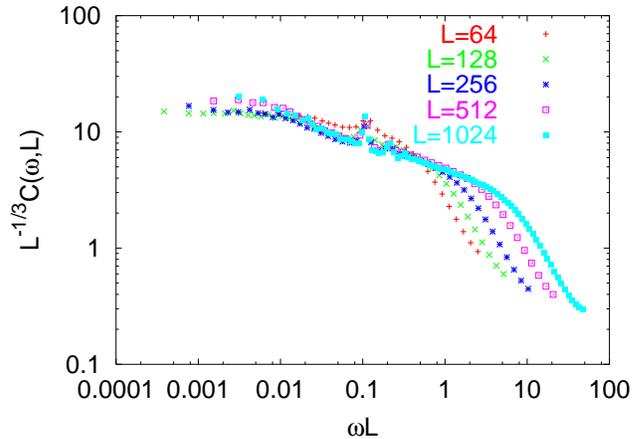}}
\caption{Scaling plot for the autocorrelation function of the energy
current, $C(\omega),$ as a function of frequency $\omega$ and system
size $L$.  The $x$-axis is $\omega L$ and the $y$ axis is $L^{-1/3}
C(\omega, L),$ in accordance with analytical predictions. All particles
have equal mass, and periodic boundary conditions are imposed in the
longitudinal direction.  The low frequency limit of the scaled functions
approaches an $L$-independent constant.}
\label{corrln}
\end{figure}

\begin{figure}
\centerline{\epsfxsize=\columnwidth \epsfbox{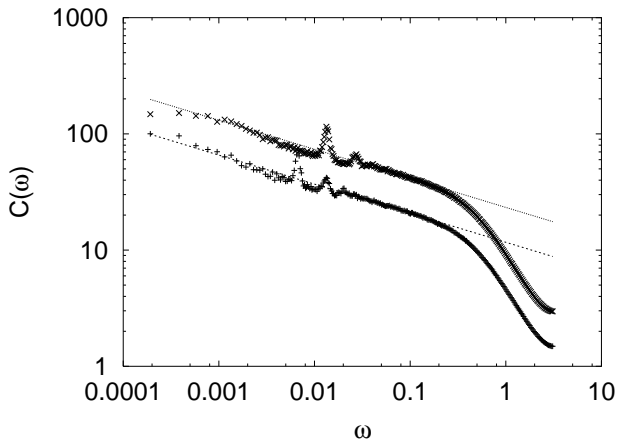}}
\caption{Unscaled plot of the autocorrelation function of the energy
current, $C(\omega),$ as a function of frequency $\omega.$ The upper
curve corresponds to a system with 512 particles and the lower curve
to a system of 1024 particles. The two curves have been shifted with
respect to each other for clarity.  All particles have equal mass, and
periodic boundary conditions are imposed in the longitudinal direction.
The straight lines correspond to a $\sim\omega^{-1/4}$ functional form.}
\label{Corrln}
\end{figure}

We have also measured the autocorrelation function for the energy current,
$C(t) = \langle J(t + t^\prime) J(t^\prime)\rangle,$ where the average
is over $t^\prime,$ since this is related to the conductivity by the
Kubo formula~\cite{Kubo}. This was done in the center of mass frame,
as necessary for the Kubo formula~\cite{bonetto}.  Figure~\ref{corrln}
shows the Fourier transform of the correlation function, $C(\omega),$
with periodic boundary conditions in the $x$ direction (without
heat reservoirs), for the case when all the particles have identical
mass. Based on the analytical prediction~\cite{ONSR}, we show $L^{-1/3}
C(\omega L)$ for a range of system sizes. The $\omega\rightarrow 0$
limit of all the curves appears to be the same, in accordance with
the numerical results we have obtained for the conductivity, although
the error bars are clearly larger due to the method of measurement. The
collapse of the curves to a single plot at low frequencies is less clear;
in fact, as shown in Figure~\ref{Corrln},
a functional form of $C(\omega, L)\sim \omega^{-1/4}$ seems to work
slightly better. The correlation function flattens out from this form at
low frequencies, at a cutoff $\omega_L$ that appears (for $L$ ranging 
from 64 to 1024, not all shown in the figure) to decrease slightly
faster than $\sim 1/L$ with $L.$ The existence of a well defined limit
$L^{-1/3}C(\omega\rightarrow 0, L)$ would require a $1/L^{4/3}$ decay for
this cutoff scale. In addition to this cutoff, there is
a clear resonance peak at $\omega\sim 1/L,$ indicating that there are 
important dynamics on the $t\sim L$ timescale.

It is not clear whether the correlation function converges to the simple
scaling form for much larger system sizes, or whether there are indeed
multiple time scales in the dynamics: $t_1(L)\sim L$ and $t_2(L) \sim
L^\beta$ with $\beta\approx 4/3.$ However, earlier work~\cite{delosRios} on
the one dimensional KPZ equation~\cite{KPZ} and work in two 
dimensions~\cite{halpin-healy}, has found it difficult to
obtain universal critical exponents from numerical simulations, even with
large system sizes. The hydrodynamic description used~\cite{ONSR}
to obtain $\alpha$ is similar to the KPZ (Burgers) equation, but with 
three equations instead of one.
In fact, $\alpha$ was correctly estimated earlier~\cite{Grass} from the 
KPZ equation. Thus the slow convergence that we see for $\alpha$ for
the pencase model --- and, to a lesser extent, for the RC
model --- may be a similar phenomenon to that seen for the KPZ equation.
Of course, even the hydrodynamic description for one-dimensional 
conduction~\cite{ONSR} does involve two time scales: for propagation 
and decay of the sound modes. 

In this paper, we have introduced a random collision model for studying
dynamics of momentum conserving one dimensional systems, in order
to obtain the scaling form of the thermal conductivity $\kappa$ as a
function of system size $L.$ This has allowed us to probe the 
scaling regime better than has been possible with previous models that
show a slower approach to the large $L$ limit and less robust behavior.
Over a wide range of length scales, we find
good agreement with the earlier analytical prediction of $\kappa\sim
L^{1/3}.$ 

We thank Abhishek Dhar and M.A. Moore for very useful discussions.


\begin{references}
\bibitem{foot1} Although the possibility that $\alpha$ could be non-universal
is also allowed by the data.
\bibitem{foot2} The range of recoil angles is of course taken to be $[0,\pi],$
{\it i.e.\/} the particle that is incident from the left in any collision
also emerges on the left.
\end{references}
\end{document}